\documentstyle[preprint,aps,pra,epsfig]{revtex}
\tighten
\begin{document}
\draft
\widetext
\preprint{version 1.01}

\title{Forbidden transitions in the helium atom}

\author{Grzegorz \L ach\thanks{present address: 
        Quantum Chemistry Laboratory, Warsaw University,
        Pasteura 1, 02-093 Warsaw, Poland}
and Krzysztof Pachucki}
\address{
Institute of Theoretical Physics, Warsaw University,
Ho\.{z}a 69, 00-681 Warsaw, Poland}
\maketitle

\begin{abstract}
Nonrelativistically forbidden, single-photon transition rates between
low lying states of the helium atom are rigorously derived within
quantum electrodynamics theory.
Equivalence of velocity and length gauges, including relativistic
corrections is explicitly
demonstrated. Numerical calculations of matrix elements are
performed with the use of high precision variational wave
functions and compared to former results.

\end{abstract}
\narrowtext The existence of nonrelativistically forbidden
transitions in helium, for example between the singlet and triplet
states, indicates the presence of relativistic effects. The
calculation of these effects in atoms or ions is a highly
nontrivial task. Depending on the magnitude of nuclear charge $Z$
one performs various approximations. Here we study light atoms, so
the expansion in the small parameter $Z\,\alpha$ is the most
appropriate. Forbidden transitions have already been studied  for 
many light atoms and especially for helium (for a review see 
\cite{john}). Historically, the first but approximate calculations 
of S-P forbidden transitions were performed by Elton in \cite{elton}.
Since the dominant part comes from $2^3P_1$ and $2^1P_1$ mixing,
he included in the calculation only these states. Drake and
Dalgarno in \cite{dd} were the first to include higher excited
states, which led to much higher precision. Moreover, Drake later
\cite{drake1} accounted for corrections to S-state wave functions.
Although these calculations were correct, there was no proof that
they are complete. As an example may serve the $2^3S_1-1^1S_0$ M1
transition. Feinberg and Sucher \cite{fein} derived an effective
operator for this transition and showed the cancellation of
electron-electron terms. However, the calculations of Drake in
\cite{drake2} were performed earlier with the implicit assumption,
that these terms are absent. In a completely different approach
based on relativistic many body perturbation theory Johnson {\em
et al.} \cite{john} and Derevianko {\em et al} \cite{der} studied
forbidden transition in both velocity and length gauge. They
pointed out the significance of negative energy states. However,
not all results were in agreement with the nonrelativistic
approach based on the Breit hamiltonian. It is the purpose of this
work to systematically derive matrix elements for forbidden
transitions in helium within quantum electrodynamics theory. 
The equivalence of length and velocity gauges for
E1 transitions, including relativistic corrections, is explicitly 
shown. With the use
of optimized numerical wave functions, the amplitudes and
transition probabilities for $2^3P_2-1^1S_0$, $2^3P_1-1^1S_0$,
$2^1P_1-2^3S_1$,  $2^3S_1-1^1S_0$, and $3^3S_1-2^3S_1$ are
calculated with high precision and compared to former results.

The nonrelativistic helium atom interacting with the electromagnetic field
is described by the Schr\"odinger-Pauli hamiltonian:
\begin{equation}
H = \frac{(\vec{p}_1-e\,\vec{A})^2}{2\,m} +
    \frac{(\vec{p}_2-e\,\vec{A})^2}{2\,m} +\frac{\alpha}{r}
-\frac{Z\,\alpha}{r_1}-\frac{Z\,\alpha}{r_2}\,.
\end{equation}
The single photon transition amplitude $T$ between two eigenstates
$\phi$ and $\psi$, in the electric dipole approximation is
\begin{equation}
T^i = \langle\phi|\frac{({p}_1+{p}_2)^i}{m} |\psi\rangle =
i\,(E_\phi-E_\psi)\,\langle\phi|({r}_1+{r}_2)^i |\psi\rangle\,,
\end{equation}
and the transition probability $\cal A$ is
\begin{equation}
{\cal A} = 2\,\alpha\,|E_\phi-E_\psi|\,T^i\,T^{*j}\,
\biggl(\delta^{ij}-\frac{{k}^i\,{k}^j}{k^2}\biggr)\,. \label{3}
\end{equation}
In the effective Hamiltonian approach relativistic corrections
enter in two ways, as corrections to the wave functions $\phi$ and $\psi$
and the correction $\delta \vec{\jmath}$ to the current $\vec{p}/m$
\begin{equation}
\vec{T} = \langle\phi|\delta \vec{j}|\psi\rangle+
\langle\phi|\frac{\vec{p}_1+\vec{p}_2}{m}\,\frac{1}{(E-H)'}\,
\delta H|\psi\rangle+
\langle\phi|\delta H\,\frac{1}{(E-H)'}\,\frac{\vec{p}_1+\vec{p}_2}{m}
|\psi\rangle\,. \label{4}
\end{equation}
The correction to the wave
function is given by the Breit hamiltonian. The part responsible for
singlet-triplet transition is
\begin{eqnarray}
\delta H = \biggl[
\frac{Z\,\alpha}{4\,m^2}\biggl(
\frac{\vec{r}_1}{r_1^3}\times\vec{p}_1-
\frac{\vec{r}_2}{r_2^3}\times\vec{p}_2\biggr)
+\frac{\alpha}{4\,m^2}\,\frac{\vec{r}}{r^3}\times(\vec{p}_1+\vec{p}_2)
\biggr]\,\frac{\vec{\sigma}_1-\vec{\sigma}_2}{2}
\equiv \vec{h}\cdot \frac{\vec{\sigma}_1-\vec{\sigma}_2}{2}\,. \label{5}
\end{eqnarray}
Corrections to the current are given by several time ordered
diagrams, shown in Fig. 1. The corresponding expression 
is calculated as follows. The first diagram is
\begin{eqnarray}
\delta \vec{\jmath}_1 &=& u^+(p')\,\vec{\alpha}\,u(p) =
\frac{1}{2\,m}\,(\vec{p}\,'+\vec{p})
-\frac{i}{2\,m}\,[(\vec{p}\,'-\vec{p})\times\vec{\sigma}]-\nonumber \\
&& \frac{1}{16\,m^3}\,(p'^2+3\,p^2)(\vec{p}+i\,\vec{p}
\times\vec{\sigma})- \frac{1}{16\,m^3}\,
(p^2+3\,p'^2)(\vec{p}\,'-i\,\vec{p}\,' \times\vec{\sigma})\,,
\label{6}
\end{eqnarray}
where $u(p)$ is a normalized plane wave solution of the free Dirac
equation. For considered transitions one may leave spin dependent
terms only. In position representation it takes a form
\begin{eqnarray}
\delta \vec{\jmath}_1 &=& \frac{i}{2\,m}\,\vec{\sigma} \times
[\vec{p},e^{i\,\vec{k}\cdot\vec{r}}] \nonumber \\ &&
+\frac{i}{16\,m^3}\,\Bigl\{
\vec{p}\times\vec{\sigma}\,e^{i\,\vec{k}\cdot\vec{r}}\,p^2+
3\,\vec{p}\times\vec{\sigma}\,p^2\,e^{i\,\vec{k}\cdot\vec{r}}-
p^2\,e^{i\,\vec{k}\cdot\vec{r}}\,\vec{p}\times\vec{\sigma}-
3\,e^{i\,\vec{k}\cdot\vec{r}}\,\vec{p}\times\vec{\sigma}\,p^2
\Bigr\}
\end{eqnarray}
The photon momentum $k$ is of order $m\,(Z\,\alpha)^2$, while $r$
is of order $(m\,Z\,\alpha)^{-1}$. This means that
$e^{i\,\vec{k}\cdot\vec{r}}$ can be expanded in powers of
$\vec{k}\cdot\vec{r}$. After adding contributions from both
electrons the $(Z\,\alpha)^2$ correction takes the form
\begin{equation}
\delta \vec{\jmath}_1 = \frac{1}{2\,m}\,(\vec{k}\cdot
\vec{r}_1)\,\vec{k}\times\vec{\sigma}_1 +
\frac{1}{2\,m}\,(\vec{k}\cdot \vec{r}_2)\,
\vec{k}\times\vec{\sigma}_2\,. \label{7}
\end{equation}
The next diagram involves one electron-positron pair and the
corresponding expression is
\begin{eqnarray}
\delta \vec{\jmath}_2 &=& -\frac{Z\,e^2}{q^2}\,\frac{1}{2\,m}\, u^+(p')
\biggl[ \vec{\alpha}\,\Lambda_-(p+q)+\Lambda_-(p'-q)\,\vec{\alpha}
\biggr]\,u(p) \nonumber \\ &=&
\frac{i}{2\,m^2}\,\frac{Z\,e^2}{q^2}\,\vec{q}\times\vec{\sigma}
\rightarrow -\frac{1}{2\,m^2}\,\frac{Z\,\alpha}{r^3}\,
\vec{r}\times\vec{\sigma}\, e^{i\,\vec{k}\cdot\vec{r}}
\,,\label{8}
\end{eqnarray}
where $\Lambda_-$ is a projection operator into the negative
energy subspace and $q$ is a momentum exchange between electron
and the nucleus. The $(Z\,\alpha)^2$ correction from both electrons
becomes
\begin{equation}
\delta \vec{\jmath}_2 =-\frac{1}{2\,m^2}\,\frac{Z\,\alpha}{r_1^3}\,
\vec{r}_1\times\vec{\sigma}_1-
\frac{1}{2\,m^2}\,\frac{Z\,\alpha}{r_2^3}\,
\vec{r}_2\times\vec{\sigma}_2\,. \label{9}
\end{equation}
The remaining diagrams involve electron-electron terms. The last two
are of higher order, so they will not be considered here. The
expression for diagram 3 can be obtained from Eq.(\ref{8}) by the
replacements $ -Z\,\alpha \rightarrow \alpha$. In this way one
obtains
\begin{equation}
\delta \vec{\jmath}_3 = \frac{1}{2\,m^2}\,\frac{\alpha}{r^3}\,
\vec{r}\times\vec{\sigma}_1\,
e^{i\,\vec{k}\cdot\vec{r_1}}+(1\leftrightarrow 2) \,,
\end{equation}
where $\vec{r}$ denotes here $\vec{r}\equiv
\vec{r}_{12}=\vec{r_1}-\vec{r}_2$. The $(Z\,\alpha)^2$ correction is
\begin{equation}
\delta \vec{\jmath}_3 =\frac{1}{2\,m^2}\,\frac{\alpha}{r^3}\,
\vec{r}\times(\vec{\sigma}_1-\vec{\sigma}_2)\,. \label{10}
\end{equation}
The expression for diagram 4 is
\begin{eqnarray}
\delta \jmath^i_4 &=&-\frac{1}{2\,m}\,\frac{e^2}{q^2}\,
\biggl(\delta^{jk}-\frac{q^j\,q^k}{q^2}\biggr)\,
u^+(p'_2)\,\alpha^k\,u(p_2) \nonumber \\ && u^+(p'_1)\biggl[
\alpha^i\,\Lambda_-(p_1+q)\,\alpha^j+\alpha^j\,\Lambda_-(p'_1-q)
\,\alpha^i\biggr]\,u(p_1) +(1\leftrightarrow 2)\,.
\end{eqnarray}
The term in the second line equals $2\,\delta^{ij}$ and
that in the first line has already appeared in Eq. (\ref{6}), so it becomes
\begin{equation}
\delta \vec{\jmath}_4 =
\frac{i}{2\,m}\,\frac{e^2}{q^2}\,\vec{q}\times\vec{\sigma}_2
+(1\leftrightarrow 2) \rightarrow \frac{1}{2\,m^2}\,\frac{\alpha}{r^3}\,
\vec{r}\times \vec{\sigma}_2\, e^{i\,\vec{k}\cdot\vec{r_1}}
+(1\leftrightarrow 2)\,.\label{12}
\end{equation}
The $(Z\,\alpha)^2$ correction is
\begin{equation}
\delta \vec{\jmath}_4 =-\frac{1}{2\,m^2}\,\frac{\alpha}{r^3}\,
\vec{r}\times(\vec{\sigma}_1-\vec{\sigma}_2)
\end{equation}
and cancels out with that from diagram 3, Eq. (\ref{10}). The
final expression for the relativistic correction to the current of
order $O(Z\,\alpha)^2$ is the sum of (\ref{7}) and (\ref{9})
\begin{eqnarray}
\delta \vec{\jmath} = \frac{1}{2\,m}\,(\vec{k}\cdot
\vec{r}_1)\,\vec{k}\times\vec{\sigma}_1 +
\frac{1}{2\,m}\,(\vec{k}\cdot \vec{r}_2)\, \vec{k}\times\vec{\sigma}_2  
-\frac{1}{2\,m^2}\,\frac{Z\,\alpha}{r_1^3}\,
\vec{r}_1\times\vec{\sigma}_1-
\frac{1}{2\,m^2}\,\frac{Z\,\alpha}{r_2^3}\,
\vec{r}_2\times\vec{\sigma}_2\,. \label{13}
\end{eqnarray}
This $\delta j$ could be also derived through the
Fouldy-Wouythusen transformation of
$\alpha^i\,e^{i\,\vec{k}\cdot\vec{r}}$, however in this way
possible electron-electron terms are omitted, which happens to be
correct for just this case. Having $\delta j$ and $\delta H$, the
transition amplitude $T^i$ in (\ref{4}) will be  transformed to
the length gauge with the use of identity
\begin{equation}
\frac{\vec{p}_1+\vec{p}_2}{m} = i\,[H,\vec{r}_1+\vec{r}_2]
\end{equation}
and the fact the terms in $T^i$ proportional to $k^i$ do not contribute
to the transition rate, as it can be seen from Eq. ({\ref{3}).
After performing simple algebraic transformations the result is
\begin{eqnarray}
T^i &=& i\,(E_\phi-E_\psi)\,
\biggl\{
\langle\phi|(r_1^i+r_2^i)\,\frac{1}{(E_\psi-H)'}\,\delta H|\psi\rangle+
\langle\phi|\delta H\,\frac{1}{(E_\psi-H)'}\,(r_1^i+r_2^i)|\psi\rangle
\biggr\} \nonumber \\ &&
+\frac{1}{2\,m}\,\epsilon^{ijk}\,\langle\phi|k^j\,T^{kl}\,k^l|\psi\rangle\,,
\label{15}
 \end{eqnarray}
where
\begin{equation}
T^{kl} = \frac{1}{2}\,\biggl[
r^k\,\frac{(\sigma_1-\sigma_2)^l}{2}+r^l\,\frac{(\sigma_1-\sigma_2)^k}{2}
-\frac{2}{3}\,\delta^{kl}\,\vec{r}\cdot
\frac{(\vec{\sigma}_1-\vec{\sigma}_2)}{2}\biggr]\,.
\end{equation}
The first term in Eq. (\ref{15}) corresponds to electric dipole,
and the second one to magnetic quadrupole transitions.
It is worth noting that for electric dipole transitions,
as given in length gauge, relativistic corrections enter
only through corrections to the hamiltonian $\delta H$.

So far, we have considered only forbidden transitions with spin
change between $S$ and $P$ states, namely $2^3P_2 \rightarrow
1^1S_0$, $2^3P_1 \rightarrow 1^1S_0$ and $2^1P_1 \rightarrow
2^3S_1$. However, even more forbidden $M1$ transitions $2^3S_1
\rightarrow 1^1S_0$ and $3^3S_1 \rightarrow 2^3S_1$ arrive at the
order $O(Z\,\alpha)^3$, so they are not described by the expression in
Eq. (\ref{13}). No second order type of terms contribute and in
the calculation of $\delta j_M$ one takes the next corresponding term
in the expansion of $e^{i\,\vec{k}\cdot\vec{r}}$ in
Eqs.(\ref{6},\ref{8},\ref{10},\ref{13})
\begin{eqnarray}
\delta \vec{\jmath}_M &=&
\frac{i}{2\,m}\,\frac{(\vec{k}\cdot\vec{r}_1)^2}{2}\,
\vec{k}\times\vec{\sigma_1}+
\frac{i}{4\,m^3}\,p_1^2\, \vec{k}\times\vec{\sigma_1}+
\frac{i}{4\,m^3}\,(\vec{k}\cdot\vec{p}_1)\,\vec{p}_1\times\vec{\sigma}_1
\nonumber \\ &&
-\frac{i}{2\,m^2}\,(\vec{k}\cdot\vec{r}_1)\,\frac{Z\,\alpha}{r_1^3}\,
\vec{r}_1\times\vec{\sigma}_1+
\frac{i}{2\,m^2}\,(\vec{k}\cdot\vec{r})\,\frac{\alpha}{r}\,
\vec{r}\times\vec{\sigma}_1+
(1 \leftrightarrow 2)\,.
\end{eqnarray}
This result agrees with the former one, obtained by Feinberg and
Sucher in \cite{fein}. For $M1$ transition between $2^3S_1$ and
$1^1S_0$ it could be further simplified to
\begin{equation}
\delta \vec{\jmath}_M = \frac{i}{m}\,\vec{k}\times
\frac{(\vec{\sigma}_1-\vec{\sigma}_2)}{2}\,
\biggl[
\frac{k^2}{12}\,(r_1^2-r_2^2)+\frac{1}{3\,m^2}\,(p_1^2-p_2^2)
-\frac{1}{6\,m}\,\biggl(\frac{Z\,\alpha}{r_1}-\frac{Z\,\alpha}{r_2}
\biggr)\biggr]\,.
\end{equation}
$k^2$ in the above can be replaced by
\begin{equation}
k^2\,r_1^2 \rightarrow [H,[H,r_1^2]] =
\frac{2}{m}\,\frac{Z\,\alpha}{r_1}-\frac{2}{m^2}\,p_1^2-
\frac{2}{m}\,\frac{\alpha}{r^3}\,\vec{r}\cdot\vec{r}_1\,,
\end{equation}
in this way one obtains for $\delta j_M$ another simple expression
\begin{equation}
\delta \vec{\jmath}_M = \frac{i}{m}\,\vec{k}\times
\frac{(\vec{\sigma}_1-\vec{\sigma}_2)}{2}\,
\biggl[
\frac{1}{6\,m^2}\,(p_1^2-p_2^2)-
\frac{1}{6\,m}\,\frac{\alpha}{r^3}\,(r_1^2-r_2^2)\,.
\biggr]
\end{equation}
The analogous expression for the $3^3S_1- 2^3S_1$ transition reads
\begin{equation}
\delta \vec{\jmath}_M = \frac{i}{m}\,\vec{k}\times
\frac{(\vec{\sigma}_1+\vec{\sigma}_2)}{2}\,
\biggl[
\frac{1}{3\,m}\,\biggl(\frac{Z\,\alpha}{r_1}+\frac{Z\,\alpha}{r_2}\biggr)-
\frac{1}{6\,m}\,\frac{\alpha}{r}
\biggr]\,.
\end{equation}

We now consider the spin algebra in the calculation of the
transition probability, as given by Eqs. (\ref{3}) and (\ref{15}).
One summs up over final states and averages out over initial states.
The appropriate formulas are:
\begin{eqnarray}
|^1S_0\rangle\langle ^1S_0| &=& |^1S\rangle\langle^1S|\,
\biggl(1-\frac{s^2}{2}\biggr)\,, \\
\frac{1}{3}\,\sum_m |^3S_1,m\rangle\langle ^3S_1,m| &=&
|^3S\rangle\langle^3S|\,\frac{s^2}{6}\,, \\
|^3P_0\rangle\langle ^3P_0| &=& |^3P^i\rangle\langle^3P^j|\,
\biggl(\delta^{ij}\,\frac{s^2}{2}-s^j\,s^i\biggr)\,, \\
\frac{1}{3}\,\sum_m|^3P_1,m\rangle\langle ^3P_1,m| &=&
|^3P^i\rangle\langle^3P^j|\,\frac{1}{2}\,s^i\,s^j\,, \\
\frac{1}{5}\,\sum_m|^3P_2,m\rangle\langle ^3P_2,m| &=&
|^3P^i\rangle\langle^3P^j|\,
\frac{1}{10}\,\biggl(2\,s^2\,\delta^{ij}-3 s^i\,s^j+2\,s^j\,s^i\biggr)\,, \\
\frac{1}{3}\,\sum_m|^1P_1,m\rangle\langle ^1P_1,m| &=&
|^1P^i\rangle\langle^3P^j|\,\delta^{ij}
\biggl(1-\frac{s^2}{2}\biggr)\,,
\end{eqnarray}
where $s = \sigma_1/2+\sigma_2/2$ and the following normalization
is utilized: $\langle P^i | P^j\rangle = \delta^{ij}/3$.
Moreover, for this calculations  one needs two formulas for spin product
\begin{eqnarray}
(\sigma_1-\sigma_2)^i\,\biggl(1-\frac{s^2}{2}\biggr)\,(\sigma_1-\sigma_2)^j
&=& 2\,\delta^{ij}s^2-4\,s^j\,s^i\,, \\
(\sigma_1-\sigma_2)^i\,s^2\,(\sigma_1-\sigma_2)^j
&=& 8\,\delta^{ij}\,\biggl(1-\frac{s^2}{2}\biggr)\,,
\end{eqnarray}
and the following set of formulas for spin traces
\begin{eqnarray}
{\rm Tr}\,s^i &=& 0\,,\\
{\rm Tr}\,s^i\,s^j &=& 2\,\delta^{ij}\,,\\
{\rm Tr}\,s^i\,s^j\,s^k &=& i\,\epsilon^{ijk}\,,\\
{\rm Tr}\,s^i\,s^j\,s^k\,s^l &=&
\delta^{ij}\,\delta^{kl} + \delta^{jk}\,\delta^{il}\,.
\end{eqnarray}
With the help of the above formulas one obtains for transition probabilities
(for simplicity we put $m=1$) the following expressions
\begin{eqnarray}
{\cal A}({^3}P_1 \rightarrow {^1}S_0) &=&
\frac{2}{9}\,\alpha\,k^3\,
\biggl|\epsilon^{ijk}\,\langle {^3}P^k|
h^i\,\frac{1}{E_P-H}(r_1+r_2)^j+
(r_1+r_2)^j\,\frac{1}{E_S-H}\,h^i|{^1}S\rangle\biggr|^2\,, \label{34}
\\
{\cal A}({^3}P_2 \rightarrow {^1}S_0) &=&
\frac{1}{30}\,\alpha\,k^5\,|\langle {^3}P^i|r^i|{^1}S\rangle|^2\,,
\\
{\cal A}({^1}P_1 \rightarrow {^3}S_1) &=&
\frac{2}{9}\,\alpha\,k^3\,\biggl|\epsilon^{ijk}\,\langle {^1}P^k|
h^i\,\frac{1}{E_P-H}(r_1+r_2)^j+
(r_1+r_2)^j\,\frac{1}{E_S-H}\,h^i|{^3}S\rangle\biggr|^2
\nonumber \\ &&+\frac{1}{18}\,\alpha\,k^5\,|\langle {^1}P^i|r^i|{^3}S\rangle|^2\,,
\\
{\cal A}({^3}S_1 \rightarrow {^1}S_0) &=&
\frac{4}{3}\,\alpha\,k^3\,\biggl|
\langle {^1}S|\frac{1}{6}\,(p_1^2-p_2^2)-
\frac{1}{6}\,\frac{\alpha}{r^3}\,(r_1^2-r_2^2)| {^3}S\rangle\biggr|^2\,,
\\
{\cal A}({^3}S_1 \rightarrow {^3}S_1) &=&
\frac{4}{3}\,\alpha\,k^3\,\biggl|
\langle {^3}S|\frac{1}{3}\,\biggl(\frac{Z\,\alpha}{r_1}+
\frac{Z\,\alpha}{r_2}\biggr)-\frac{1}{6}\,\frac{\alpha}{r} |{^3}S\rangle
\biggr|^2\,,
\end{eqnarray}
where $k=|\Delta E|$, and $h^i$ is defined by Eq. (\ref{5}).
It is worth noting that $^1P_1 \rightarrow {}^3S_1$ is not only $E1$ transition
but also $M2$, which has not yet been recognized in the literature.

Once transition probabilities are expressed in terms of
matrix elements between nonrelativistic wave functions,
they can be calculated numerically with high precision.
In the numerical calculation we follow an approach developed by
Korobov \cite{kor}. The wave function is expressed in terms of exponentials
\begin{eqnarray}
\phi_S &=& \sum_i c_i\bigl[
e^{-\alpha_i\,r_1 -\beta_i\,r_2-\gamma_i\,r}\mp
(r_1\leftrightarrow r_2)\bigr]\,, \\
\vec{\phi}_P &=& \sum_i c_i\bigl[
\vec{r}_1\,e^{-\alpha_i\,r_1 -\beta_i\,r_2-\gamma_i\,r}\mp
(r_1\leftrightarrow r_2)\bigr]\,,\\
\vec{\phi}_{P+} &=& \sum_i c_i\,\vec{r}_1\times\vec{r}_2\,\bigl[
e^{-\alpha_i\,r_1 -\beta_i\,r_2-\gamma_i\,r}\mp
(r_1\leftrightarrow r_2)\bigr]\,.
\end{eqnarray}
The parameters $\alpha_i,\beta_i,\gamma_i$ are chosen randomly
between some minimal and maximal values, which were found by
minimization of energy of a specified state. The maximal dimension
of this basis set was 600. Lower values were used for checking
convergence. The advantage of this basis set is simplicity of
matrix elements, which are expressed in terms of integral
\begin{equation}
\frac{1}{16\,\pi^2}\int d^3 r_1\,d^3 r_2\,
\frac{e^{-\alpha\,r_1 -\beta\,r_2-\gamma\,r}}{r_1\,r_2\,r} =
\frac{1}{(\alpha+\beta)\,(\beta+\gamma)\,(\gamma+\alpha)}\,.
\end{equation}
For some more singular matrix elements an additional integral
with respect to corresponding parameters has to be performed.
The disadvantage of this basis set is the necessity of
using quadruple precision for $N>100$. Moreover,
the second order terms require more careful tuning
of parameters due to the singularity of $\delta H$ and large mixing
of $2^3P_1$ and $2^1P_1$ states.
These, which involve odd parity intermediate $P$-states, are much larger
than those which involve even parity $P$-states, by approximately
three orders of magnitude. It is due to the fact that energies
of even parity $P$ states lie beyond the ionization level.
Most often, these small  second order terms were neglected in the former
calculations. However, they are not neglected here.
Our numerical results for forbidden transitions
between low lying states are presented in Table I.

In the comparison with former work we start with
the $M1$ transition $2^3S_1\rightarrow 1^1S_0$. This transition
was measured by Moos and Woodworth in \cite{moos} with the result
${\cal A} = 1.10(33)\,10^{-4}\,{\rm s}^{-1}$ and
Gordon Berry from Notre Dame is currently preparing a more precise measurement.
The first (correct) theoretical result obtained by Drake  in \cite{drake2} 
$1.272\cdot 10^{-4}\,{\rm s}^{-1}$ is in agreement with the experimental value.
However, as pointed out by Feinberg and Sucher in \cite{fein}, Drake has not
considered electron-electron terms, which happened to cancel 
out for this transition. Later, Johnson {\em et al} \cite{john} used 
RMBPT to calculate forbidden transitions for
any helium-like ions and obtained a result for $Z=2$,
which is $1.266\cdot 10^{-4}\,{\rm s}^{-1}$. It differs slightly
from the result obtained here $1.272426\cdot 10^{-4}\,{\rm s}^{-1}$,
due to inclusion in \cite{john} of some higher order terms,
while electron correlations were not well accounted for.
Moreover there are unknown radiative corrections 
and exchange type of diagrams of order $\alpha/(2\,\pi)$, the last two in Fig. (1),
to any of these transitions. Therefore only first 3 digits
are physically significant. Numerical results are presented with higher precision
for the purpose of comparison with former results.
Next the M1 transition $3^3S_1\rightarrow 2^3S_1$ rate was obtained only by
Derevianko {\em et al} in \cite{der}. Their result $1.17\,10^{-8}\,{\rm s}^{-1}$,
disagrees with ours, $6.484690\,10^{-9}\,{\rm s}^{-1}$.
The reason of this discrepancy is left unexplained.
It may indicate the loss of accuracy of RMBPT due to strong numerical cancellation.
This discrepancy does not have experimental impact since this rate is too small
for $Z=2$ to be measured. However, calculations should be verified for
higher $Z$, where this transition rate grows with $Z^{10}$ and becomes measurable
at some value of $Z$.
The next considered transition is M2: $2^3P_2\rightarrow 1^1S_0$.
It was first obtained by Drake \cite{drake1}: ${\cal A} = 0.327\,{\rm s}^{-1}$, and later by
Johnson {\em et al} \cite{john}  ${\cal A} = 0.3271\,{\rm s}^{-1}$,
in agreement with our result ${\cal A} = 0.3270326\,{\rm s}^{-1}$.
The calculation of the intercombination E1 transition $2^3P_1\rightarrow 1^1S_0$
was little more elaborate, since it involves infinite summation
over intermediate states. In former works the second term in Eq. (\ref{34})
involving even parity P-states was neglected. Indeed, calculations
show it is smaller than 1\%. The first complete result by Drake \cite{drake1} is
$ {\cal A} = 176.4\,{\rm s}^{-1}$. RMBPT calculations of Johnson {\em et al} \cite{john}
including negative energy states is $ {\cal A} = 175.7\,{\rm s}^{-1}$ and
our result $ {\cal A} = 177.5771\,{\rm s}^{-1}$ agrees within 1\%.
The last transition $2^1P_1\rightarrow 2^3S_1$ is a sum of E1 and M2.
The result $ {\cal A} = 1.55\,{\rm s}^{-1}$ obtained by Drake includes only E1 transition.
Our result is $ {\cal A} =1.548945\,{\rm s}^{-1}$ and the magnetic transition
happened to be negligible $0.000019$ due to small energy splitting.

In summary, we have presented a rigorous derivation of rates for nonrelativistically
forbidden transitions. We demonstrated equivalence of length and velocity gauges
including relativistic correction for forbidden transitions. 
We confirmed the commonly used fact
that in the length gauge relativistic corrections enters only through
corrections to wave function as given by Breit hamiltonian.
We verified that M2  $2^1P_1\rightarrow 2^3S_1$ transition is much smaller
than E1, which was implicitly assumed in former works.
Our numerical calculations using simple exponential functions
confirmed former results with the exception
of $3^3S_1\rightarrow 2^3S_1$ transition,
where our result is approximately twice smaller than of \cite{der}.

\section*{ACKNOWLEDGMENTS}
The work of K.P. was supported by Polish Committee for Scientific Research
under Contract No. 2P03B 057 18. The work of G.{\L}. was done
in partial fulfillment of the requirements for the M.Sc. degree.

\begin{figure}
\centerline{\epsfig{figure=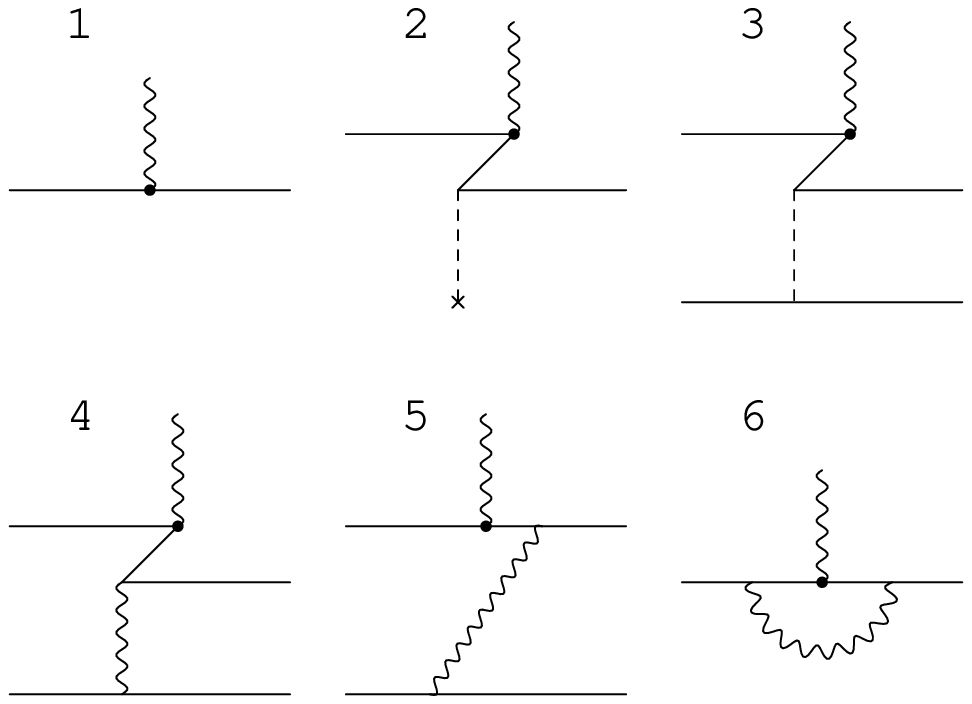,width=5in}}
 \caption{Time ordered diagrams for corrections to the current.
 Dashed line is a Coulomb photon, the wavy line is the transverse photon. }
\label{fig1}
\end{figure}

\begin{table}
\caption{Transition rates in helium in units s$^{-1}$, $[n] \equiv 10^n$}
\label{tab1}
\begin{tabular}{rll}
transition  &$\Delta E$ in atomic units &rate ${\cal A}$ \\
\tableline
E1+M2: $2^1P_1\rightarrow 2^3S_1$   & 0.0513862917& 1.548945   \\
E1: $2^3P_1\rightarrow 1^1S_0$   & 0.7705606863& 1.775771[2]   \\
M2: $2^3P_2\rightarrow 1^1S_0$   & 0.7705606863& 3.270326[-1]  \\
M1: $2^3S_1\rightarrow 1^1S_0$   & 0.7284949988& 1.272426[-4]  \\
M1: $3^3S_1\rightarrow 2^3S_1$   & 0.1065403108& 6.484690[-9]  \\
\end{tabular}
\end{table}
\end{document}